# Luminescent Defects in Single-Walled Carbon Nanotubes for Applications

*Jana Zaumseil**

Prof. J. Zaumseil
Institute for Physical Chemistry, Universität Heidelberg, D-69120 Heidelberg, Germany
E-mail: zaumseil@uni-heidelberg.de

Semiconducting single-walled carbon nanotubes show extraordinary electronic and optical properties, such as high charge carrier mobilities and diameter-dependent near-infrared photoluminescence. The introduction of $sp^3$ defects in the carbon lattice of these nanotubes creates new electronic states that result in even further red-shifted photoluminescence with longer lifetimes and higher photoluminescence yield. These luminescent defects or organic color centers can be tuned chemically by controlling the precise binding configuration and the electrostatic properties of the attached substituents. This review covers the basic photophysics of luminescent $sp^3$ defects, synthetic methods for their controlled formation and discusses their application as near-infrared single-photon emitters at room temperature, in electroluminescent devices, as versatile optical sensors, and as fluorophores for bioimaging and potential super-resolution microscopy.

# 1. Introduction

Single-walled carbon nanotubes (SWCNT) have many extraordinary mechanical, electronic and optical properties that make them an interesting material for basic research but also for many applications. They are generally described as seamlessly rolled-up sheets of graphene with roll-up angles and diameters that are determined by the chiral vector ($C_h$) and identified by the indices (n,m).[1-2] Their one-dimensional bandstructure can be derived from that of graphene when imposing the periodic boundary conditions created by the respective chiral vector. Due to the many different possible diameters (~ 0.6 – 2.5 nm) and chiral angles (0 – 30°), there are well over a hundred unique SWCNTs (i.e., different chiralities) that are produced by the various growth techniques. They can be either metallic or semiconducting and each have characteristic optical transitions.[3-4] For most applications only semiconducting nanotubes and ideally only one (n,m) species (monochiral) are desired. This technological need has pushed the search for more controlled and selective growth of SWCNTs[5-6] and even more importantly fueled the development of post-growth purification and sorting of the different nanotube types and chiralities. Various, quite different techniques, such as gel-chromatography,[7-9] aqueous two-phase separation (ATPE)[10-12] and selective polymer-wrapping,[13-17] now make it possible to produce and work with sufficiently large amounts of not only purely semiconducting nanotubes of a certain diameter range but also monochiral nanotubes and even enantiomerically pure samples.[18-19]

This high degree of purification and hence the availability of nanotubes as a material with reproducible properties has been critical for applications in electronics[20-21] and energy conversion,[22-23] as well as imaging[24-25] and sensing[26-27] based on the near-infrared photoluminescence (PL) of SWCNTs and its modulation. However, an important paradigm shift occurred when it became clear that defects in the sp² carbon lattice of nanotubes are not always detrimental to the photoluminescence yield, but can indeed lead to new electronic states with highly interesting and useful optical properties.[28-30] These specific defects, which are variably named sp$^3$-defects, luminescent defects, organic color centers or quantum defects[31-35] have been at the heart of many recent studies on carbon nanotubes and promise a wide range of potential applications from single-photon emission to in-vivo super-resolution imaging. This review will briefly introduce the underlying photophysics and properties of these defects, peruse recent progress in their controlled creation and discuss the various potential applications in detail.

## 2. Photophysics of luminescent defects in SWCNTs

Semiconducting SWCNTs have rich photophysics that have been covered extensively in various reviews.[36-38] This review will, however, mainly focus on the impact of defects on the excitonic emission from semiconducting carbon nanotubes in the near infrared (NIR). Excitons (bound electron-hole pairs) in SWCNTs have very large binding energies (several 100 meV) due to the reduced dielectric screening.[39] The degeneracy of the K and K' valleys of the underlying graphene lattice creates four distinct singly-excited configurations KK, K'K', KK' or K'K with the electron (hole) residing in the K(K) valley in addition to one singlet and three triplet states for each of these valley configurations.[40-42] Hence there are in total 16 possible excitons. However, only one of them is bright (the zero-momentum, odd parity singlet exciton), resulting in the typical $E_{11}$ photoluminescence peaks of semiconducting nanotubes with minimal Stokes shift (few meV) and narrow linewidth.[43] Excitons in SWCNTs have very high diffusion constants (1-10 cm$^2$ s$^{-1}$)[44] and are thus able to sample large parts of the nanotube (several hundred nanometers) during their lifetime (few ps). This fast diffusion is one of the reasons for the low photoluminescence (PL) yield (at best few %) of nanotubes even in purified samples (without metallic nanotubes),[16, 45] as the highly mobile excitons are very likely to find quenching sites (see **Figure 1a**), e.g. the ends of the nanotube, unintentional structural defects, charge carriers, etc. They are thus also very sensitive to the immediate environment of the nanotube (e.g. solvent, surfactant coating, polymer-wrapping or adsorbed molecules)[46-48] and the PL yield drops dramatically to well below 0.5% for very short nanotubes (<100 nm)[44, 49-50] and barely detectable signals for ultrashort nanotubes (<50 nm).[51]

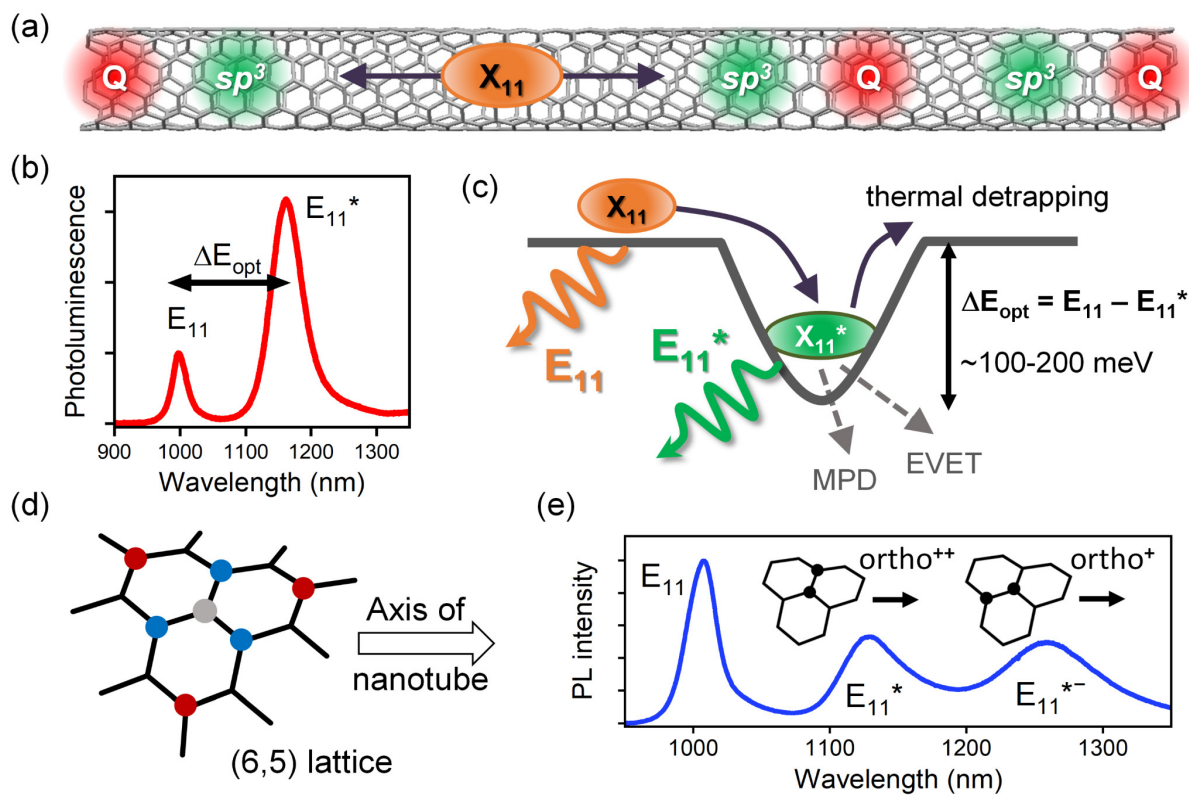

**Figure 1.** (a) Schematic illustration of a nanotube with a mobile exciton ($X_{11}$), several quenching sites (Q), e.g., nanotube ends and luminescent $sp^3$ defects. Representative photoluminescence spectrum of a dispersion of (6,5) nanotubes functionalized via diazonium chemistry to create luminescent defects with corresponding $E_{11}^*$ emission. The optical trap depth $\Delta E_{opt}$ is indicated. (c) Model of exciton dynamics at $sp^3$ defects as proposed by He et al.[52] including non-radiative loss channels: multiphonon decay (MPD), electronic-to-vibrational energy transfer to solvent (EVET), and thermal detrapping. (d) Binding configurations of substituents to a (6,5) nanotube lattice fragment. The creation of the initial $sp^3$ carbon atom (gray) is followed by a binding event at ortho (blue) or para (red) positions, all of which create different defect states. (e) PL spectrum of (6,5) SWCNTs functionalized with two types of $sp^3$ defects with different ortho binding configurations that result in $E_{11}^*$ and $E_{11}^{*-}$ emission.

The introduction of $sp^3$ defects through controlled chemical functionalization, that is, converting an $sp^2$ hybridized carbon to an $sp^3$ carbon, perturbs the electronic structure of the nanotube, breaks the local symmetry and creates new optically active excitonic states.[33] The new $E_{11}$* state is energetically below the bright $E_{11}$ state and thus any mobile exciton encountering this defect will be localized. It becomes trapped in the lower energy state. A new red-shifted $E_{11}$* emission appears while the intrinsic $E_{11}$ emission is reduced (see **Figure 1b**). The typical trap depth (~ 100 – 250 meV)[29] is deep enough to prevent escape of the exciton at room temperature and thus also diffusion to other non-radiative defect sites or the nanotube ends. The only non-radiative relaxation mechanisms for localized excitons are multiphonon decay (MPD), electronic-to-vibrational energy transfer to the surrounding solvent (EVET), and some thermal detrapping as schematically shown in **Figure 1c** based on the model by He et al.[52]

The inhibited diffusion to quenching sites and non-radiative decay of excitons there can substantially improve the total PL yield of nanotubes functionalized with a limited number of luminescent defects.[29, 53] While the precise quantification of the defect density in an ensemble sample is difficult, it is usually possible to correlate the ratio of the intensities of the $E_{11}$ to $E_{11}$* emission (**Figure 1b**) with the D/G ratio from resonant Raman spectra of the same sample as a metric for comparison.[29, 53-54] The estimated defect concentrations are actually very low with about 5-10 defects per micrometer for maximum PL yield. Indeed, the positions of single defects and thus localized emission spots have been spatially and spectroscopically resolved for individual long and short nanotubes.[51, 55-56] While such a low number of defects is barely detectable in absorption spectra of dispersions or thin films, they dominate the emission spectra as the majority of mobile excitons are very efficiently and quickly funneled to the defect sites. At higher defect densities, the carbon lattice and hence the electronic structure of the nanotubes becomes too perturbed, which results in quenching and an overall decrease in total PL yield. The maximum enhancement factor for optimized defect densities depends on the nanotube diameter[29] but also on its length and initial PL yield. For example, very short SWCNTs (<50 nm) that barely show any PL due to strong quenching at the nanotube ends, become bright and easily detectable when luminescent $sp^3$ defects are introduced.[51] Long (>1 µm) polymer-wrapped (6,5) nanotubes, however, which already exhibit good PL yield (~ 2 %), only show very moderate enhancement (factor 2).[53] Other studies have reported PL yield enhancement of 5 to 10 times for (6,5) SWCNTs in aqueous dispersion.[29] Unfortunately, even with such strong enhancement, nanotubes (especially in dispersion or thin films) are still outperformed by near-infrared emitting core-shell quantum dots, such as InAs/CdSe with ~ 30% of absolute PL yield at 1080 nm.[57]

One important advantage of the low absorption of and enhanced emission from a small number of defects in SWCNTs is the lack of reabsorption of the emitted light,[58] which can be a problem for highly concentrated nanotube dispersions or films, due to the small Stokes shift.[59] This effect is also beneficial for radiative pumping of exciton-polaritons in strongly-coupled microcavities with functionalized (6,5) carbon nanotubes without changes in the polariton branch structure.[60]

Time-correlated single-photon counting (TCSPC) measurements on nanotubes with $sp^3$ defects have shown that the photoluminescence lifetime of the $E_{11}$* defect emission is much longer (100-500 ps) than that of the mobile excitons (few ps). This effect is due to the limited non-radiative decay paths (MPD, EVET, see **Figure 1c**) and inhibited thermal detrapping at room temperature for localized excitons. Note that the mobile excitons - created by optical excitation along the entire nanotube - populate the luminescent defect states within a few ps.[61] The long lifetime of the trapped excitons means that even moderate excitation densities lead to a saturation of the defect sites. Thus, the $E_{11}/E_{11}$* intensity ratio becomes highly pump power-dependent,[53, 62] which makes the comparison of defect densities in experimental studies with different excitation sources (lamp, pulsed or continuous wave laser) very difficult.

Although the (6,5) nanotubes (diameter 0.76 nm, bandgap ~ 1.27 eV) are the SWCNT species most often used for functionalization due to their availability and easy purification, luminescent defects in various SWCNT chiralities with different diameters were investigated. An inverse correlation between the nanotube diameter and the optical trap depth ($\Delta E_{opt} = E_{11} - E_{11}^*$) of an sp$^3$ defect was found,[29, 63] which also affects the PL lifetime of the defect emission (longer for deeper traps). The trap depth can be tuned to some degree by the substituents that are directly attached to the sp$^3$-carbon and their electrostatic properties. More electron-withdrawing moieties lead to a deeper trap depth and longer PL lifetimes as well as larger PL yield enhancement, while an electron-donating character results in more shallow traps and shorter PL lifetimes.[29, 53, 64-65] However, the differences between the maximum and minimum values achieved by such electrostatic variations are in the range of 40 meV or less in energy and about 50 ps in lifetime.

Much more impact can be assigned to the binding configuration of the luminescent defects. Although we typically speak of "one sp$^3$ defect" with a specific molecular substituent, actually two sp$^2$ carbon atoms must be converted to sp$^3$-hybridized carbon (the second possibly saturated with a molecular fragment from the solvent) to restore charge neutrality and a closed-shell system. The position of the two sp$^3$ carbon atoms with respect to each other determines the energy of the defect state and thus the wavelength of emission. Due to the low symmetry of the majority of semiconducting SWCNTs (except the achiral zig-zag nanotubes) there are three different possible ortho positions (blue) and three different para positions (red) in relation to the first sp$^3$ carbon as shown in **Figure 1d**. The presence of these different defect binding configurations in an ensemble of functionalized nanotubes creates spectral diversity that can be resolved by single nanotube PL spectroscopy at cryogenic temperatures.[66]

Time-dependent density functional theory (TD-DFT) calculations have helped to assign the different emission features to specific binding configurations for selected nanotube chiralities.[66-69] Based on the current understanding, ortho arrangements are preferred and the most common binding configurations in chiral SWCNTs are ortho$^{++}$ and ortho$^+$ (the labels vary)[68], which give rise to the $E_{11}*$ and more red-shifted $E_{11}*^-$ emission as visualized in **Figure 1e** for (6,5) nanotubes. This dependence of the different emission features on these rather subtle differences between the binding configurations in chiral nanotubes was further confirmed by the functionalization of achiral (11,0) nanotubes, which only showed $E_{11}*^-$ emission.[68]

As discussed above the optical trap depth correlates with the photoluminescence lifetime and that of the $E_{11}*^-$ emission typically reaches 500-600 ps for $\Delta E_{opt} \sim 250$ meV compared to 150-200 ps and $\Delta E_{opt} = 150 -180$ meV for $E_{11}*$ of (6,5) nanotubes. Both values decrease for nanotubes with larger diameters.[54, 70]

Note that unfortunately there is no unified nomenclature for the defect type and binding configuration of sp$^3$-functionalized carbon nanotubes at this point. A number of different labels (including $E_{11}^-$, $E_{11}^{2}*$ etc.) have been used in the literature and unambiguous assignment of a broad emission peak to a specific binding configuration is not always clear. For the purpose of this review, the first prominent defect emission with the smaller optical trap depth will always be referred to as $E_{11}*$ and that associated with a deeper trap-depth and more red-shifted emission as $E_{11}*^-$.

## 3. Synthetic methods to introduce luminescent defects

The methods to introduce sp$^3$ defects to carbon nanotubes are diverse and depend strongly on the nanotube sample, i.e., in aqueous dispersion with surfactants such as sodium dodecyl sulfate,[29] polymer-wrapped in organic solvents,[53] deposited on a substrate[71] or freestanding as-grown nanotubes.[72] Successful functionalization has been shown for all of these examples, although the most common methods use aqueous dispersions and only recently also dispersions in organic solvents. **Figure 2** shows a limited selection of reactions that create luminescent sp$^3$ defects. Many more variations have been reported. Several reviews have provided detailed overviews on the different chemical reactions,[32, 34-35] such that this review will focus on a few generalized concepts and more recent examples. Note that it is also possible to covalently functionalize carbon nanotubes without affecting the conjugation of the carbon lattice as shown by Setaro et al. using a [2+1] reaction with azidodichlorotriazine.[73] This method can also be used to attach further functional groups for sensing and imaging[74-75] without new red-shifted emission peaks.

The first intentional creation and explicit observation of luminescent defects was made for ozone-treated SWCNTs in aqueous dispersion and termed O-doping.[28] Ozonation probably produces oxygen defects via ozonide intermediates that decompose to epoxides or ethers (**Figure 2a**).[28, 76] Recently, a number of simple methods to produce O-doped nanotubes with defect emission were reported, such as the reaction with hypochlorite under UV irradiation,[77] photo-oxidation with lipid hydroperoxides[78] or even just deposition of oxides such as $Al_2O_3$ or $SiO_2$ onto nanotubes by electron beam evaporation.[71]

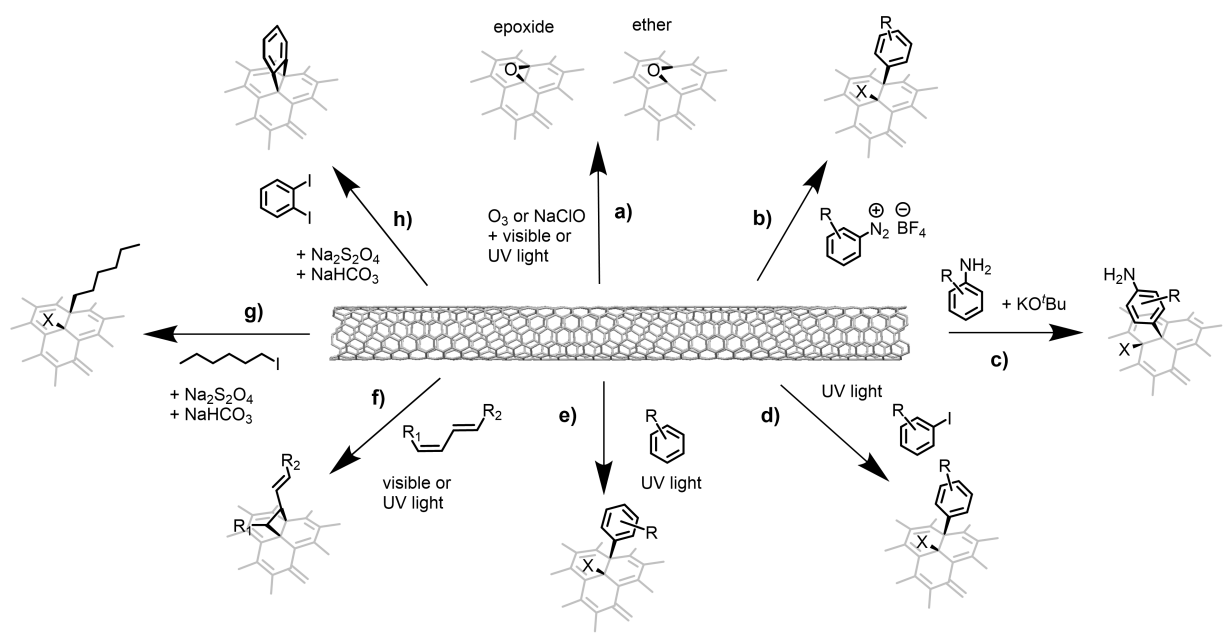

**Figure 2.** Selection of possible chemical reactions to create luminescent defects in carbon nanotubes and suggested structures of likely binding configurations. (a) Ozonation/O-doping with ether and epoxy groups;[28, 30, 77] (b) reaction with pre-formed aryldiazonium salts;[29, 53] (c) nucleophilic addition with 2-haloanilines in the presence of a strong organic base;[54] (d) UV light-induced activation of aryl halides[79] or (e) nitroarenes and aminoarenes; [80] (f) [2π + 2π]

photocycloaddition of enones;[81] (g) reductive coupling with alkyl iodides or (h) arylene diiodides.[82]

However, the covalent attachment of aryl or alkyl groups to SWCNTs has received much more attention than O-doping due to the chemical tunability of the substituents and often brighter defect emission. In particular, aryl diazonium chemistry (**Figure 2b**) is a popular and very versatile functionalization method. During the reaction of SWCNTs with the diazonium salt the $sp^2$ lattice is probably attacked by an aryl radical that is formed either in a Gomberg-Bachmann reaction[83] or through a single-electron transfer from the SWCNT to the diazonium salt followed by release of nitrogen. The final defect density can be easily controlled by the diazonium salt concentration and the reaction time as the diazonium reaction is relatively slow and often takes several hours at room temperature, although it can be accelerated by irradiation with visible light.[84-85] In addition, a one-pot process in which the diazonium compound is generated in situ from an aromatic amine is possible and has been demonstrated.[86] Overall a large variety of functional moieties have been attached to $sp^3$ defects through diazonium chemistry and thus enabled their various applications, e.g., as optical sensors (see **section 6**).
Some functional groups are not compatible with diazonium chemistry or the resulting compounds have a limited shelf life. An alternative for otherwise inert aryl compounds is activation by UV light. As schematically shown in **Figure 2d**, aryl iodides react with nanotubes under UV irradiation.[79] Amino- and nitroarenes also introduce luminescent defects (**Figure 2e**) under such conditions, although the molecular structure of the attached groups and the mechanism are not yet clear.[80]Alkyl groups can be attached to nanotubes using a Billups-Birch type reduction[87-89] or with alkyl iodides activated by the reducing agent sodium dithionite (**see Figure 2g**)[82, 90] The latter route is very versatile and can also be applied to create divalent defects with aryl diiodides (**Figure 2h**). A comprehensive overview of the various possible reaction schemes was recently given by Janas.[35]

The vast majority of reactions with carbon nanotubes to intentionally create luminescent defects has been carried out in aqueous dispersions and with nanotubes stabilized by typical surfactants such as alkyl sulfates, cholates or DNA. This is partially due to the fact that nanotubes purified by gel-chromatography or ATPE are already in aqueous dispersions and often need to be biocompatible for application in biological samples. However, polymer-sorted semiconducting nanotubes are usually dispersed in organic solvents such as toluene or xylene.[91] They are highly attractive for (opto)electronic applications due to the high purity (semiconducting and even monochiral) that can be reached by selective polymer-wrapping with minimal effort.[13, 16, 92-95] SWCNT dispersions in organic solvents with higher viscosity and higher vapor pressure than water are also more suitable for uniform film formation by spin-coating or printing for functional devices on a larger scale.[96-98] Organic solvents further enable a wider range of possible chemical reactions and reaction conditions (anhydrous, oxygen-free) that might be employed to create specific $sp^3$ defects.

First demonstrations of the direct functionalization of polymer-wrapped nanotubes with the goal of a controlled luminescent defect density used either dip-doping of pre-deposited nanotubes with very reactive diazonium salts[70] or in situ generation of aryl diazonium from 4-nitroaniline, which required elevated temperatures and inert conditions.[99-100] A scalable, simple and very controllable method to facilitate the reaction of aryl diazonium salts with polymer-wrapped nanotubes is the addition of a phase-transfer agent (a crown ether) and a polar co-solvent (acetonitrile) to the dispersion. Berger et al. demonstrated that monochiral (6,5) nanotubes that were sorted and wrapped with a polyfluorene-bipyridine copolymer in toluene could be functionalized this way with different aryl-substituents at room temperature, showing $E_{11}$* emission and enhanced PL yield after work-up.[53] Other recent examples of the controlled introduction of $sp^3$-defects to polymer-wrapped nanotubes in organic solvents include the reaction with benzoyl peroxide[101] and 2-haloanilines.[54] Others are certainly possible and should further expand the options for $sp^3$ functionalization of chirality-sorted nanotubes.

As outlined above the precise binding configuration, i.e., the position of the sp$^3$ carbons with respect to each other, is crucial for the emission properties of the final defect state and emission wavelength. Being able to determine what defects are formed by applying suitable synthetic methods is an important step toward reproducible samples and thus applications. The controlled creation of the defect configurations (ortho$^+$, see **Figure 1e**) with more red-shifted emission ($E_{11}$*$^-$) has gained increased attention recently, as they have deeper optical trap-depths, are better suitable for room-temperature single-photon emission and emit closer to the desired wavelength range for telecommunication and the second biological window. For example, the $E_{11}$*$^-$ emission of (6,5) nanotubes is around 1280 nm (compared to ~ 1160 nm for $E_{11}$*), while that of (10,3) nanotubes is at 1550 nm (C-Band).[70] Previously, nanotubes that showed $E_{11}$*$^-$ emission were only a minority byproduct of the main functionalization process, e.g., aryl diazonium chemistry, which predominantly leads to $E_{11}$* defects (i.e., ortho$^{++}$ configuration). However, various functionalization strategies can produce at least as many $E_{11}$*$^-$ as $E_{11}$* defects in one sample. Shiraki et al. first reported an additional, more red-shifted emission peak for (6,5) nanotubes functionalized with a bisdiazonium compound that promoted proximal covalent modification.[102] The idea of using substituents with two adjacent functional groups (bidentate) for reactions with the nanotube lattice to obtain $E_{11}$*$^-$ emission was further demonstrated for the reductive alkylation with dibromoalkanes.[103] Other methods that promote significant but not only $E_{11}$*$^-$ emission are the photoreaction with aromatic compounds such as iodoaniline under argon[80] or the [2π + 2π] photocycloaddition of enones[81] (see **Figure 2e** and **2f**). An elegant synthetic method to create exclusively sp$^3$ defects with ortho$^+$ configuration for $E_{11}$*$^-$ emission is the reaction with 2-haloanilines in the dark and in the presence of a strong organic base such as potassium tert-butoxide (KO$^t$Bu, see **Figure 2c**).[54] The nucleophilic intermediates formed by deprotonation should preferentially attack C–C bonds with large π-orbital misalignment, i.e., those in axial direction of the nanotube. Extra illumination with UV light adds a possible radical reaction path toward defects with $E_{11}*$ emission and thus allows for tuning the ratio of the two emission peaks.

Many more reaction schemes could be tested for the creation of different types of luminescent defects, especially those methods that were previously used for the covalent functionalization of nanotubes[104] without considering low defect concentrations or photoluminescence. Both spatial and spectral control over the created defects and their emission wavelengths are desired for future applications in addition to a larger variety of attached functional groups that can either serve as starting points for further modifications or exhibit interesting properties (e.g., spin) themselves (see **sections 6** and **7**).

## 4. Single-photon emitters

One of the many exciting properties of semiconducting SWCNTs is their potential for single-photon emission due to exciton localization even without any intentional functionalization and in seemingly homogeneous environments.[105-106] Early experiments had demonstrated photon antibunching (i.e., vanishing probability of the emission of more than one photon per excitation) in nanotubes at cryogenic temperatures[107] and thus introduced them as strong contenders for single-photon emitters in secure quantum communication and even quantum computing.[108] Carbon nanotubes are particularly attractive for these applications for several reasons. Their emission already occurs in the near-infrared, i.e., close to the typical telecommunication bands (O-band, C-band), and it can be excited not just optically but also electrically in individual nanotubes (see electroluminescence, **section 5**). Drawbacks however are the low quantum yield of SWCNTs, their propensity to spectral diffusion and blinking,[109-111] and with few exceptions (where either exciton-exciton annihilation or non-covalent patterning of the energy landscape were employed)[112-115] the need for cryogenic temperatures to achieve localization. The introduction of covalent and luminescent $sp^3$ defects alleviates at least the latter limitation.

The first demonstration of room-temperature single-photon emission from carbon nanotubes was provided by Ma et al. who created luminescent defects in individual, surfactant-wrapped (6,5) nanotubes by electron-beam deposition of $SiO_2$ on and around them.[71, 116] These defects were speculated to be the result of either epoxy or ether groups, similar to those created by ozonation chemistry,[76] with deep enough trap potentials (100 meV or more) to sufficiently localize excitons even at higher temperatures. Both $E_{11}^*$ and $E_{11}^{*-}$ emission were observed and Hanbury-Brown-Twiss (HBT) experiments were carried out on a large number of single nanotubes, of which many showed photon antibunching with a zero-delay second-order photon correlation $g^{(2)}(0) < 0.5$ at temperatures up to 298 K.

With the more controlled chemical functionalization of various nanotubes chiralities and the tunable introduction of specific defects using aryl diazonium chemistry (see **section 3**), reliable photon antibunching for SWCNTs at room temperature with high single-photon purities, high emission rates of 10–100 MHz, and wavelengths close to the telecommunication bands could be achieved. **Figures 3a-c** show an example of a polymer-wrapped (6,5) nanotube functionalized with a dichloroaryl substituent and exhibiting $E_{11}^{*-}$ emission at 1280 nm with $g^{(2)}(0) = 0.01$ at room temperature as reported by He et al.[70] The long fluorescence lifetime (>500 ps) of the defect emission is indicative of the large optical trap depth as thermal detrapping and non-radiative decay of mobile excitons at quenching sites or tube ends is prevented. Following these seminal reports, single-photon emission at room temperature was demonstrated for a number of nanotubes functionalized in different ways and with different substituents.[54, 68, 90] Theoretical limits for single-photon purity of $sp^3$-functionalized nanotubes were discussed based on Monte Carlo simulations.[117]

For coherent single-photon emitters with indistinguishability of the emitted photons, the ratio of total dephasing time (limited by the frequency of exciton scattering events) to twice the spontaneous emission lifetime must be increased. Various cavity-enhancement methods can be employed to speed-up radiative decay and shorten the emission lifetime in comparison to the dephasing time, thus also increasing single-photon emission rates from $sp^3$ defects. The Purcell effect of a photonic or plasmonic cavity in the weak coupling regime can increase the radiative decay rate (i.e., reduce emission lifetime) of carbon nanotubes substantially.[118-119] One approach was demonstrated by Ishii et al. who integrated aryl functionalized (6,5) nanotubes on a two-dimensional silicon photonic crystal microcavity as shown in **Figure 3d**.[120] Defect emission that was coupled to the cavity mode showed very narrow linewidth (few nm, given by the cavity quality factor), 50 times higher photoluminescence intensity and a 30% shorter emission lifetime (see **Figure 3e**) compared to uncoupled nanotubes with the same defects. The single-photon purity at room temperature was still very high with $g^{(2)}(0) = 0.1$ and stable for a wide range of excitation powers. This example also demonstrated that functionalized nanotubes can be integrated as single-photon emitters in silicon photonics.

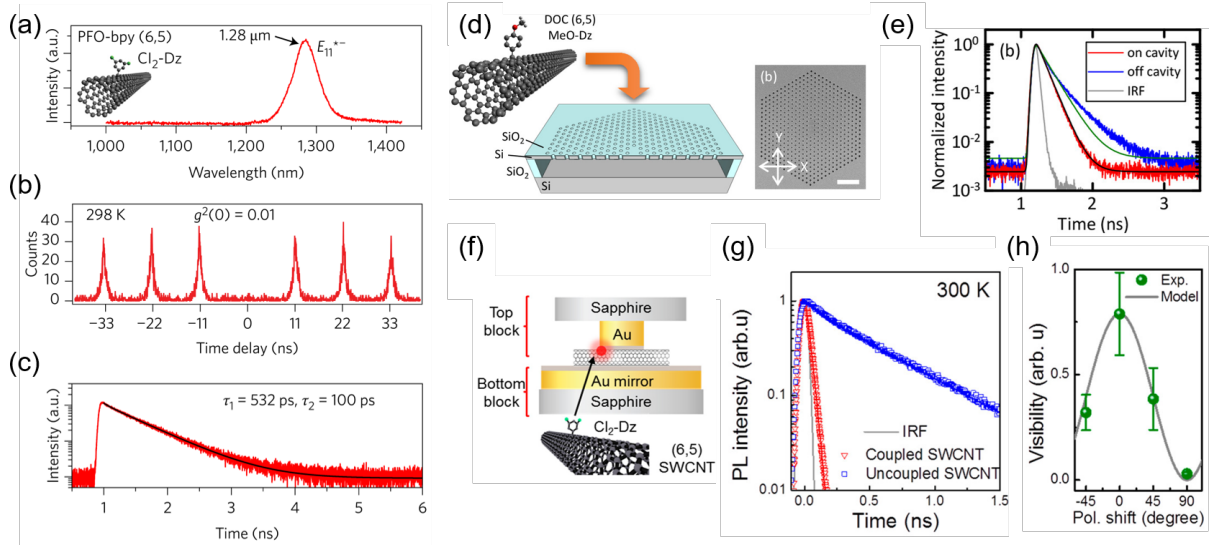

**Figure 3.** (a) $E_{11}^{*-}$ defect emission (at 1280 nm) from a single polymer-sorted (6,5) nanotube functionalized with a dichloroaryl group. (b) Corresponding second-order photon correlation $g^{(2)}(t)$ with high 99% single-photon purity and (c) biexponential photoluminescence decay traces. Reproduced and adapted with permission.[70] Copyright 2017, Springer Nature. (d) Schematic layout of two-dimensional photonic crystal with functionalized (6,5) nanotubes and (e) corresponding photoluminescence decay traces for coupled and uncoupled SWCNTs showing reduction of emission lifetime by the Purcell effect. Reproduced and adapted with permission.[120] Copyright 2018, American Chemical Society. (f) Illustration of nanotubes with luminescent defects in a plasmonic cavity with substantially reduced emission lifetime as shown by time-resolved photoluminescence (g). Two-photon interference visibility (at 4 K) as a function of the polarization shift between photons in the two arms of a Hong-Ou-Mandel interferometer. Reproduced and adapted with permission.[121] Copyright 2019, American Chemical Society.

An even stronger effect was observed for plasmonic gap mode nanocavities with a much smaller mode volume. For $sp^3$ functionalized nanotubes (with $E_{11}^*$ and $E_{11}^{*-}$ defects) sandwiched between a gold nanocube array and an $Al_2O_3$ covered gold mirror (see **Figure 3f**), Luo et al. found defect emission lifetimes of only 11 ps compared to 588 ps for uncoupled nanotubes (**Figure 3g**). The emission intensity of the coupled nanotubes was about 52 times higher than for uncoupled SWCNTs and a high single-photon purity of 99 % was maintained at 300 K.[121]

To also lengthen the dephasing time with the goal of photon indistinguishability, the samples were cooled to 4 K and two-photon interference (TPI) visibility was measured with a Hong-Ou-Mandel interferometer. The TPI visibility reached up to 0.79 (**Figure 3h**), which indicated high photon indistinguishability and coherent single-photon output.

Overall these encouraging reports highlight the real potential of sp$^3$ functionalized nanotubes as single photon emitters in the telecommunication wavelength range. However, they still need to be transferred successfully to practical photonic platforms (e.g., waveguides) and, most importantly, integrated in electroluminescent devices for real applications.

## 5. Electroluminescent devices

Shortly after semiconducting single-walled carbon nanotubes were found to be excellent materials for field-effect transistors with high electron and hole mobilities,[122] electroluminescence (EL) from a single nanotube was demonstrated by Misewich et al in 2003.[123] This pioneering experiment opened the prospect of nanoscale light sources in the technologically important telecommunication range. It was quickly followed by a number of examples of EL resulting from electron-hole recombination[124-126] or impact excitation[127-128] in carbon nanotubes. Indeed, the first truly ambipolar light-emitting transistors were based on individual semiconducting SWCNTs before thin film transistors with other luminescent semiconductors such as conjugated polymers,[129-130] organic single crystals[131-132] and colloidal quantum dots[133] followed.

The general concept of light-emitting transistors has been discussed in detail in a couple of recent reviews.[134-135] Carbon nanotubes are special in this regard as they are intrinsically ambipolar, the injection barriers for both holes and electrons are fairly low and their carrier mobilities are much higher than those of other typical emitter materials. The formation of accumulation layers of holes and electrons within the channel depending on the applied gate and drain voltages is easily accomplished with nanotubes as long as electron traps such as water/oxygen[136] are removed. SWCNTs are thus highly suitable for ambipolar light-emitting transistors, not just as individual nanotubes[124, 137-139] but also as networks of sorted semiconducting nanotubes.[140-143] However, the low photoluminescence and even lower electroluminescence efficiencies of SWCNTs remain serious obstacles for any application. The low electroluminescence efficiencies in field-effect devices might be partially due to the interaction of the nanotubes with polar device substrates,[144-145] Auger quenching by excess charge carriers[143, 146-147] or non-radiative energy transfer and decay in dense networks.[148] However, even light-emitting diodes with semiconducting nanotubes embedded in a polymer matrix show very low external quantum efficiencies (<0.02 %).[149]

The introduction of luminescent defects may offer two important advantages: electrically pumped single photon emission as discussed in **section 4** and increased external quantum efficiency of electroluminescent devices. As mentioned in **section 2** the photoluminescence quantum yield of sp$^3$-functionalized nanotubes can be increased substantially compared to the pristine nanotubes for optimized defect concentrations.[29, 51, 53-54] While the exact enhancement factors vary depending on the type of functionalization and starting material (e.g., chirality and length of nanotubes), the effect should be translatable to electroluminescent devices. However, carrier mobility is an important factor in light-emitting transistors as it determines the maximum current densities. Hence, the impact of luminescent defects on charge transport must be considered as well.

Usually, defects and trap states of any kind are deemed detrimental to charge transport in individual SWCNTs or networks and are thus avoided as much as possible. Early studies of the effect of covalent defects introduced to individual nanotubes (metallic or semiconducting) via diazonium chemistry also indicated substantial losses in conductivity.[150-151] The first luminescent devices with intentionally sp$^3$-functionalized nanotubes used a simple two-electrode structure (no gate) and relied on impact excitation to electrically generate excitons.[152]

Relatively long (6,5) nanotubes (> 2 µm) functionalized with 4-nitroaryl groups were dropcast on gold electrodes (distance 5 µm) while an alternating electric field was applied to form semi-aligned networks. Photoluminescence spectra from the network showed $E_{11}$ and defect emission. The same and even further red-shifted emission features were observed as spotty electroluminescence from the channel region for applied electric fields above 0.015 MV/cm. The emission intensity increased exponentially with current, which is consistent with exciton formation by impact excitation.[127, 153]

Defect electroluminescence resulting from electron-hole recombination in ambipolar light-emitting field-effect transistors was shown by Zorn et al.[154] Top-gate/bottom-contact transistors with dense networks of semiconducting nanotubes and a PMMA/HfO$_x$ gate dielectric (**see Figure 4a-c**) had been previously shown to enable highly reproducible ambipolar charge transport[155-157] and the formation of a very controlled recombination and emission zone within the channel.[158-160] Networks of polymer-sorted (6,5) SWCNTs with covalent bromoaryl defects showed decreasing hole and electron mobilities with increasing defect density (quantified by the D/G Raman ratios), indicating their role as charge traps. However, even at relatively high degrees of functionalization (when the total quantum yield drops again) the transistors remained fully functional with reasonable carrier mobilities (about one third to a quarter of the values for pristine nanotubes, see **Figure 4d,e**). The recorded EL spectra were similar to the photoluminescence spectra from the channel with $E_{11}$, $E_{11}^*$ and $E_{11}^{*-}$ emission peaks, although with a somewhat larger share of $E_{11}^{*-}$. The emission intensity increased almost linearly with drain current, while the spectral distribution changed such that the defect emission became slightly less dominant compared to the $E_{11}$ peak (see **Figure 4f**). Overall these experiments showed that the sp$^3$ defects acted as shallow charge traps while excitons were either directly created at the defects or funneled to them as in the case of optical excitation (PL). Unfortunately, no clear enhancement of the external quantum efficiency could be observed for the functionalized networks compared to the pristine (6,5) SWCNTs.

These recent examples clearly show that electrical excitation of emission from luminescent defects is possible and no less efficient than emission from pristine nanotubes. The impact of defects on charge transport is significant but does not completely prevent normal device operation. It remains to be seen whether the optimization of defect density per nanotube, number of functionalized nanotubes in a network or type of defect can ultimately result in sufficiently bright defect emission for practical electroluminescent devices.

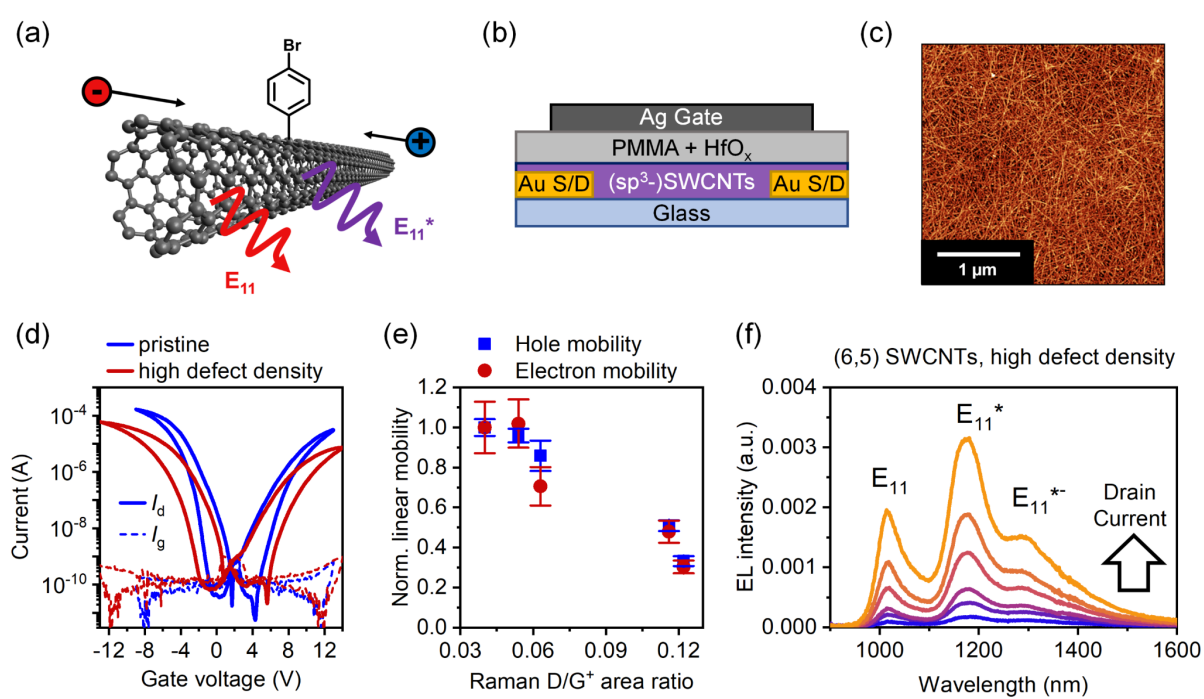

**Figure 4.** Schematic illustration of electroluminescence based on electron-hole recombination (a) and device structure of a top-gate/bottom-contact transistor (b) with a dense network (AFM image) of functionalized (6,5) nanotubes (c). Transfer characteristics for pristine versus highly functionalized (6,5) SWCNT network (d) and normalized hole and electron mobilities depending on defect concentration as quantified by $D/G^+$ Raman signal ratio (e). EL spectra (f) of functionalized nanotubes (with $E_{11}$, $E_{11}^*$ and $E_{11}^{*-}$ emission) for increasing drain currents. Reproduced and adapted under the terms of the Creative Commons BY-NC-ND 4.0 license.[154] Copyright 2021, American Chemical Society.

## 6. Optical Sensors

The substituents of luminescent sp³ defects in carbon nanotubes can be used for further conjugation with functional groups that may acts as sensors for various analytes or environmental conditions, by altering the wavelength or intensity of the defect emission selectively. Such optical sensors - in contrast to sensors based on changing the electrical resistance of nanotubes or nanotube networks[161] - are particularly interesting for micro- or nanoscale environments where local information is required and macroscopic techniques and bulky devices fail. Nanotubes that are non-covalently functionalized with DNA, lipids, peptides or other polymers have already been applied extensively as fluorescent nanoprobes[24, 162-166] with spatial and spectral resolution.[167] The advantage of luminescent sp³ defects is the separation of the general interaction of the nanotube with its environment leading to changes in the $E_{11}$ emission and a specific interaction with the functional group on the defect, thus modifying only the $E_{11}*$ or $E_{11}*^-$ emission. A shift or change of intensity of one emission peak in relation to another is much more easily reproduced and analyzed than variations of a single emission feature that may vary for a number of reasons.

A variety of optical nanoprobes based on luminescent defects has been demonstrated over the last few years. A first example was the pH sensitive emission from sp³ defects with *N,N*-diethyl-4-aminobenzene groups.[168] Depending on the pH of the aqueous dispersion and thus change of the amino moiety from protonated (low pH, electron withdrawing) to deprotonated (high pH, electron donating), the defect emission peak shifted to lower wavelengths, i.e., higher energies, by about 18 meV, while the $E_{11}$ peak remained constant (see **Figure 5a,b**). Comprehensive studies of sp³ defects with various aryl-substituents had shown a linear correlation of the optical trap depth with the Hammett substituent constants.[64] Consequently, any change of the electron-withdrawing or -donating properties of a functional group attached to a defect, e.g., by a specific binding event (molecular recognition), can be employed for sensing when using the $E_{11}$ emission as a reference.

This concept was indeed explored by Shiraki et al. who functionalized (6,5)-rich CoMoCAT SWCNTs with phenylboronic acid groups. The formation of a boronate with a diol compound led to a blue-shift of the defect emission peak due to stronger electron donation as demonstrated by the addition of different saccharides such as D-fructose and D-glucose (see **Figure 5c,d**).[169] Similarly when using a *N*-phenylaza-15-crown-5-ether as a functional group on the sp$^3$-defect, a clear redshift of the $E_{11}*$ emission was observed after adding Ag$^+$ ions to the aqueous dispersion. The selective binding of the silver cation by the azacrown ether decreases the electron-density at the nitrogen atom due to the coordination of the electron lone pair to the metal.[170] As there is a large variety of ionophores for the specific binding of different metal ions, it should be possible to create a range of cation selective optical sensors based on sp$^3$-functionalized SWCNTs.

General properties of the environment of a dispersed nanotube (solvent etc.) can be inferred from the different defect emission intensities in relation to the $E_{11}$ peak as shown by Liu et al.[171] They employed (6,5) SWCNTs functionalized with dimethylaniline groups yielding two defect emission peaks ($E_{11}*$ and $E_{11}*^-$) and investigated the change of intensity ratios (not the peak positions) depending on the dynamic viscosity of the dispersion. They assumed that the rotation of the dimethylaniline group (see **Figure 5e**) constitutes a non-radiative loss channel for the defect emission, which should be reduced in higher viscosity solvents. Hence, they added glycerol to the aqueous dispersion of the functionalized nanotubes to change the viscosity over three orders of magnitudes and test this hypothesis. Indeed, the ratio of defect to mobile exciton emission increased for higher viscosities, in particular the more red-shifted $E_{11}*^-$ emission gained in intensity compared to $E_{11}$ (see **Figure 5f,g**). The peak intensity ratios could thus be used as a ratiometric measurement of the local viscosity. Furthermore, it would be interesting to correlate these spectroscopic viscosity measurements with those based on the diffusion coefficients that can be extracted from tracking the Brownian motion of a single nanotubes.[25, 172]

The notion of using the ratio of the $E_{11}$ to defect ($E_{11}^*$ or $E_{11}^{*-}$) emission intensity as a measure for additional non-radiative decay channels close to the sp$^3$ defects was further explored by Berger et al. with polymer-wrapped (6,5) SWCNTs that were covalently functionalized with the stable perchlorotriphenylmethyl (PTM) radical via diazonium chemistry.[173] The neutral open shell system of the PTM substituent at the sp$^3$ defect resulted in lower defect emission as well as shorter fluorescence lifetimes compared to nanotubes with the same number of defects but with a bromoaryl substituent. Again, the emission wavelengths were very similar. Furthermore, the direct transformation of the PTM radical via UV light irradiation and subsequent reaction to create a closed shell system immediately increased the defect emission and restored the longer fluorescence lifetime. These experimental observations strongly suggest a radical induced quenching mechanism, with photoinduced electron transfer (PET) and radical enhanced intersystem crossing (EISC)[174] being the most likely candidates (see **Figure 5h,i**). These examples highlight the sensitivity of the defect emission efficiency to the precise nature of the substituent and the surrounding medium.

Overall, the targeted functionalization of sp$^3$ defects with moieties that enable specific binding through molecular recognition or with groups that are in other ways sensitive to external stimuli and analytes creates a large variety of potential optical probes. Both the relative shift of the defect emission peak and the relative change of defect emission intensity compared to the $E_{11}$ emission can be used as quantitative metrics. Importantly they offer a higher reliability and selectivity than peak position and intensity changes of only the $E_{11}$ emission of non-covalently functionalized nanotubes.

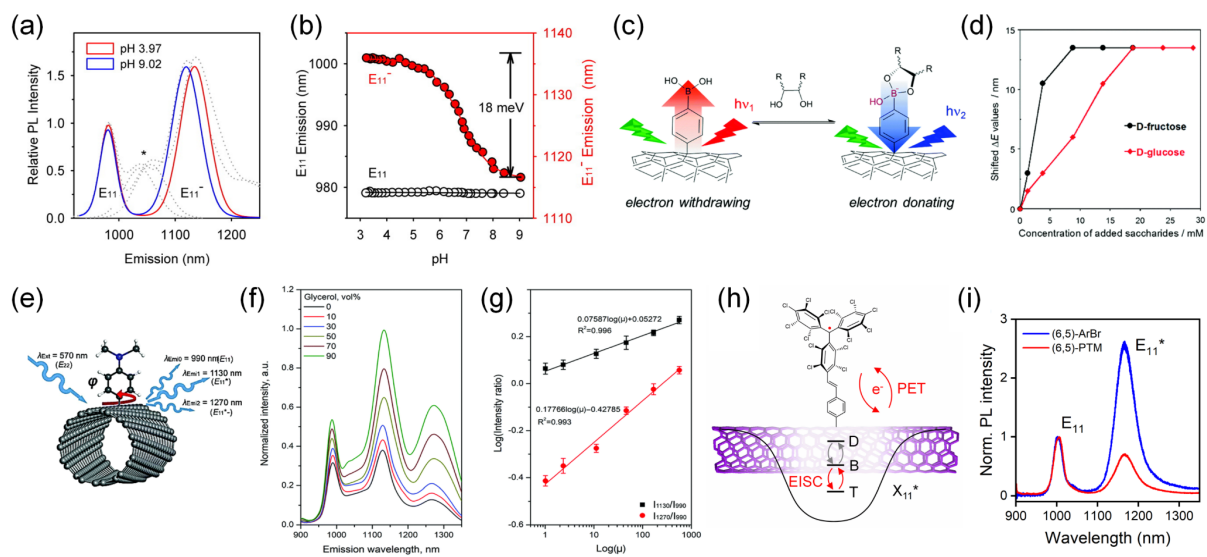

**Figure 5.** Defect emission of N,N-diethyl-4-aminobenzene functionalized (6,5)-SWCNTs. The defect emission (here labelled $E_{11}^-$) shifts from 1117 nm to 1136 nm with decreasing pH whereas $E_{11}$ remains unchanged. Reproduced with permission.[168] Copyright 2015, American Chemical Society. (c) Boronate formation with a diol compound changes the electron withdrawing/donating properties of the boronic acid group attached to a defect leading to a blue-shift of the defect emission peak as demonstrated for different saccharides (d). Reproduced with permission.[169] Copyright 2016, Royal Society of Chemistry. (e) (6,5) SWCNTs functionalized with a dimethylaniline group resulting in two defect emission peaks $E_{11}^*$ and $E_{11}^{*-}$. (f) Corresponding PL spectra for different glycerol/water mixtures and (g) dependence of the intensity ratios of $E_{11}^*/E_{11}$ and $E_{11}^{*-}/E_{11}$ on dynamic viscosity (μ) of the solution. Reproduced with permission.[171] Copyright 2020, Royal Society of Chemistry. (h) Interactions of the unpaired spin of a PTM radical with the $sp^3$ defect of a (6,5) nanotube including photoinduced electron transfer (PET) and radical enhanced intersystem crossing (EISC) leading to lower $E_{11}^*$ emission compared to a reference sample with the same number of defects but with bromoaryl (ArBr) substituents (i). Reproduced under the terms of the Creative Commons BY-NC-ND 4.0 license.[173] Copyright 2021, American Chemical Society.

## 7. Bioimaging

Single-walled carbon nanotubes were first successfully employed as emitters for in-vivo near-infrared fluorescence imaging by the group of Hongjie Dai in 2009, visualizing the blood vessels of a living mouse through the skin and later even through the skull.[175-176] Nanotubes are exceptionally suitable for this application based on their emission in the 2nd biological window (1000-1350 nm, NIR-II) while they can be excited in the 1st biological window (700-900 nm). The NIR-II range is particularly attractive due to low light scattering and absorption by biological tissue thus enabling deeper light penetration and better spatial resolution. While other NIR emitters (e.g., quantum dots and organic fluorophores)[57, 177] are available, nanotubes combine a number of advantageous properties. They can be made highly bio-compatible through non-covalent functionalization, e.g., with DNA[178] and their photostability compared to organic fluorophores aids imaging over long periods of time. However, the low PL yield especially of short SWCNTs is a major obstacle for applications. The introduction of luminescent $sp^3$ defects with even further red-shifted emission can make even very short nanotubes (<50 nm) bright enough for detection[51] and may facilitate improved super-resolution microscopy of single nanotubes moving within living tissue or cells.[25, 179] Mandal et al. reported successful imaging of single (6,5) SWCNTs with luminescent defects in live brain tissue with ultralow excitation doses.[180] These nanotubes were made bio-compatible by phospholipid-polyethylene glycol and showed bright emission at 1160 nm ($E_{11}^*$ emission) when excited at the $E_{11}$ transition (985 nm), thus combining strong absorption and emission with low background fluorescence (see **Figure 6a,b**). This detection approach also avoided any tissue damage as the excitation intensity of 100 W/cm$^2$ was an order of magnitude lower than that required for detecting individual pristine nanotubes.

As shown above, the brightness of single nanotubes as fluorophores is significantly increased by introducing luminescent defects. The photostability and photoluminescence yield of nanotubes can be further improved by coating them with crosslinked polymer layers.[110, 181] An elegant example of creating $sp^3$ defects while simultaneously forming a stable polymer-coating was recently shown by Nagai et al. with the radical polymerization of N-isopropyl acrylamide to PNIPAM within the surfactant micelle of (6,5) SWCNTs dispersed in water.[182] The resulting functionalized nanotubes were stable in aqueous dispersion even without surfactant and showed red-shifted defect emission with overall much greater brightness than previous polymer-coated nanotubes without luminescent defects. With further modifications of the polymer-coating[183] they might become ideal near-infrared emitters for bioimaging.

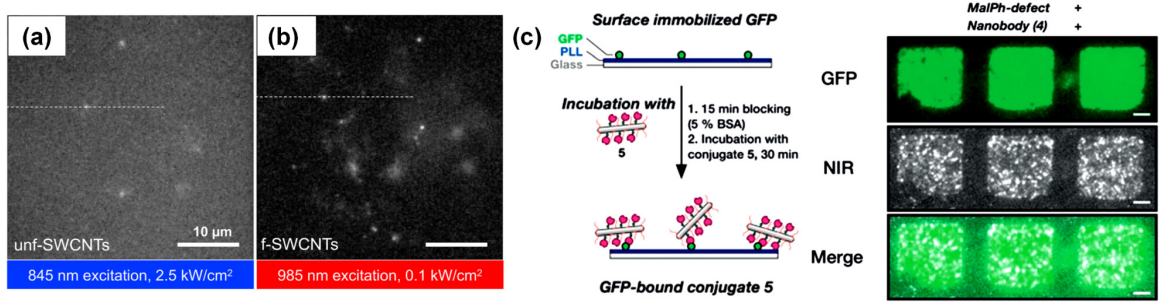

**Figure 6.** Fluorescence images of (a) unfunctionalized (6,5) SWCNTs (845 nm excitation; 2.5 kW/cm$^2$) and (b) functionalized SWCNTs (985 nm excitation; 0.1 kW/cm$^2$) collected from depths >15 μm inside live brain slices. Reproduced under the terms of the Creative Commons Attribution 4.0 International License.[180] Copyright 2020, Springer Nature. (c) Pattern of green fluorescent protein (GFP) on a glass surface incubated with SWCNT functionalized with GFP-binding protein. The colocalization shows retained functionality of the nanobody even after covalent conjugation to the SWCNT. Scale bars are 5 μm. Reproduced under the terms of the Creative Commons CC BY license.[184] Copyright 2020, Wiley-VCH.

In addition to just acting as NIR emitters for imaging, sp$^3$-defects also facilitate the covalent attachment of additional fluorophores or biomolecules such as proteins and peptides to nanotubes without PL quenching. Mann et al. demonstrated the sp$^3$-functionalization of nanotubes with of (*N*-maleimido)phenyl groups and fluorenylmethoxycarbonyl (Fmoc)-protected l-phenylalanine groups via diazonium chemistry.[184] These functional groups were subsequently used to attach nanobodies that can bind the green fluorescent protein (GFP) as a proof of principle. The successful binding of the GFP to the SWCNTs was confirmed by imaging and colocalization of the near-infrared emission of the nanotubes and green emission from the GFP (see **Figure 6c**). In addition, it was possible to directly grow peptide chains from the nanotubes and use these for further interactions. Such combination of luminescent sp$^3$-defects and biomolecules could be expanded much further and should enable new applications in bioimaging and biosensing with nanotubes.

# 8. Future Directions and Conclusion

Since the initial realization that the controlled introduction of certain defects to the sp$^2$-carbon lattice of semiconducting SWCNTs can create new electronic states with red-shifted photoluminescence substantial progress has been made. The molecular origin of the emissive states and their dependence on nanotube chirality, binding configuration and chemical nature of the substituents is largely understood. It is possible to control the type of defects and hence defect emission via a variety of synthetic methods and the obtained functionalized nanotubes can be applied directly as single-photon emitters or in electroluminescent devices. They can be further modified to serve as selective and highly responsive optical nanoprobes for imaging and sensing especially in biological samples (live tissue or cell cultures). However, the inhomogeneity of the defect distribution remains a big challenge. While it might be possible to reproduce an average defect density for an ensemble of nanotubes in dispersion, the number and position of defects on each individual nanotube remains statistical. The deterministic localization of luminescent defects is desired especially for the goal of electrically pumped single-photon emitters. On-site reaction with the help of either photolithographically opened access to a nanotube[151] or optically induced reactions[185] might be an option to overcome this challenge. Furthermore, the continued exploration of novel functional groups, e.g., with attached fluorescent probes, ionophores for metal ion sensing, open shell systems with spin states, single molecular magnets or plasmonic nanoparticles, and their impact on the optical and electronic properties of the sp$^3$-defects and the rest of the nanotube should provide a large space of opportunities for this newest incarnation of carbon nanotubes as an exciting and useful nanomaterial.

**Acknowledgements**

J.Z. has received funding from the European Research Council (ERC) under the European Union's Horizon 2020 research and innovation programme (Grant agreement no. 817494 "TRIFECTs").